\documentstyle[prb,aps,eqsecnum,epsf,preprint,floats]{revtex}

\begin{document}

\tightenlines

\title{Spontaneous Magnetization of the O(3) Ferromagnet at Low Temperatures}

\author{Christoph P. Hofmann}

\address{Department of Physics, University of California at San Diego,
9500 Gilman Drive, La Jolla, California 92093}

\date{June 2001}

\maketitle

\begin{abstract}
\noindent
We investigate the low-temperature behavior of ferromagnets with a spontaneously broken
symmetry O(3) $\to$ O(2). The analysis is performed within the perspective of
nonrelativistic effective Lagrangians, where the dynamics of the system is formulated in
terms of Goldstone bosons. Unlike in a Lorentz-invariant framework (chiral perturbation
theory), where loop graphs are suppressed by two powers of momentum, loops involving
ferromagnetic spin waves are suppressed by three momentum powers. The leading
coefficients of the low-temperature expansion for the partition function are calculated
up to order $p^{10}$. In agreement with Dyson's pioneering microscopic analysis of the
cubic ferromagnet, we find that, in the spontaneous magnetization, the magnon-magnon
interaction starts manifesting itself only at order $T^4$. The striking difference with
respect to the low-temperature properties of the O(3) antiferromagnet is discussed from a
unified point of view, relying on the effective Lagrangian technique.
\end{abstract}


\section{Introduction}
\label{Intro}

In the following presentation, our interest is devoted to the low-temperature behavior of
ferromagnets, which exhibit a spontaneously broken internal symmetry O(3) $\to$ O(2).
This system has been widely studied in condensed matter physics and it is not our
intention to contribute to its detailed microscopic understanding. Rather, we want to
analyze several low-energy phenomena from a unified point of view, relying on the method
of nonrelativistic effective Lagrangians. We try to understand how the symmetries,
inherent in the underlying theory, manifest themselves in the partition function at low
temperatures. The complex microscopic description of the system is taken into account
only through a phenomenological parametrization, which, in the effective Lagrangian,
emerges in the form of a few coupling constants.

Nevertheless, let us first consider the Heisenberg model, which describes the ferromagnet
on a {\it microscopic} level. There, the exchange Hamiltonian ${\cal H}_0$,
\begin{equation}
\label{Heisenberg}
{\cal H}_0 \ = \ - \, J \, \sum_{n.n.} {\vec S}_m \! \cdot {\vec S}_n \, , \ \ \ 
J = const \, ,
\end{equation}
formulates the dynamics in terms of spin operators ${\vec S}_m$, attached to lattice
sites $m$. Note that the summation only extends over nearest neighbors and that the
isotropic interaction is assumed to be identical for any two adjacent lattice sites. For
positive values of the exchange integral $J$, the above expression leads to an adequate
low-energy description of the ferromagnet.
 
The interaction between a constant magnetic field ${\vec H} = (0,0,H), H > 0$, and the
spin degrees of freedom is taken into account through the Zeeman term. In the
corresponding extension of the Heisenberg model,\cite{footnote1}
\begin{equation}
\label{Zeeman}
{\cal H} \ = \ {\cal H}_0 \, - \, {\mu}{\sum_n}{\vec S}_n \! \cdot {\vec H} \, , 
\end{equation}
the magnetic field is coupled to the vector of the total spin.

Within this model, Dyson evaluated the low-temperature expansion for the partition
function of a ferromagnet for all three types of cubic lattices.\cite{Dyson} In
particular, for the most prominent order parameter of the ferromagnet, the spontaneous
magnetization, he obtained the following series:
\begin{equation}
\label{MagnetizationTemp}
\Sigma(T) \, / \, \Sigma(0) \ = \ 1 \, - \, {\alpha}_0 T^{\frac{3}{2}} \, - \,
{\alpha}_1 T^{\frac{5}{2}} \, - \, {\alpha}_2 T^{\frac{7}{2}} \, - \, {\alpha}_3 T^4
+ \ {\cal O} (T^{\frac{9}{2}}) \, .
\end{equation}
The $T^{3/2}$-term, referred to as Bloch's law, corresponds to free magnons. The next two
terms involving the coefficients ${\alpha}_1$ and ${\alpha}_2$ are related to the shape
of the dispersion curve -- or the discreteness of the lattice -- and describe free
magnons as well. Remarkably, the spin-wave interaction starts manifesting itself only at
order $T^4$.

In order to obtain this result, Dyson had to set up a highly involved mathematical
machinery. It is our goal to rederive this series within the effective Lagrangian
framework and to understand Dyson's result by exclusively considering the symmetries of
the theory. In order to do so, we will first have to establish the momentum power
counting scheme for this nonrelativistic system, which represents the very basis for a
systematic expansion of quantities of physical interest. As we will see, it is quite
different from the power counting scheme in Lorentz-invariant effective theories (chiral
perturbation theory).

Another aspect of the present work concerns the comparison of the low-energy properties
of O(3) ferromagnets and O(3) antiferromagnets. Within the framework of nonrelativistic
effective Lagrangians, we will, e.g., be able to understand the striking differences in
the low-temperature expansions for the order parameters: the spontaneous magnetization
for a ferromagnet and the staggered magnetization for an antiferromagnet, respectively.

In the effective Lagrangian perspective, the excitations near the ground state, the spin
waves or magnons in the present case, are interpreted as Goldstone bosons resulting from
a spontaneously broken internal symmetry. Indeed, the Heisenberg Hamiltonian
(\ref{Heisenberg}) is invariant under a simultaneous rotation of the spin variables
described by the symmetry group G = O(3), whereas the ground state of a ferromagnet
breaks this symmetry spontaneously down to H = O(2): all the spins are aligned in one
specific direction, giving rise to a nonzero spontaneous magnetization.

Whenever a physical system exhibits spontaneous symmetry breaking and the corresponding
Goldstone bosons represent the only low-energy excitations without energy gap, we do have
a very powerful means at our disposal to analyze its low-energy structure: effective
Lagrangians. The method was originally developed in connection with Lorentz-invariant
field theories (chiral perturbation theory),\cite{Weinberg junior,Coleman Callan,Li
Pagels,Weinberg seminal,Gasser Leutwyler QCD} admitting a low-energy analysis of the
strong interaction described by quantum chromodynamics (QCD). The effective Lagrangian
technique has also proven to be very useful in the investigation of other relativistic
systems where Goldstone bosons occur, see, e.g.,
Refs.\cite{Hasenfratz Leutwyler,Hasenfratz Niedermayer 1991,Hasenfratz Niedermayer 1993}.
The method has been extended to finite temperature,\cite{Gasser Leutwyler} allowing for a
systematic low-temperature analysis of the partition function.

In condensed matter physics, spontaneous symmetry breaking is a common phenomenon and
effective field theory methods are widely used in this domain. Only recently, however,
has chiral perturbation theory been extended to such nonrelativistic
systems,\cite{Leutwyler NRD,Roman Soto,Roman Soto Canted} demonstrating its applicability
to solid state physics as well. The method is based on effective Lagrangians which
exploit the symmetry properties of the underlying theory, i.e., the Heisenberg model in
the present case, and permits a systematic low-energy expansion of quantities of physical
interest in powers of inverse wavelength. The few applications related to systems
exhibiting collective magnetic behavior that appeared in the literature so far concern
spin-wave scattering processes,\cite{Hofmann spin wave} spin-wave mediated
non-reciprocal effects in antiferromagnets,\cite{Roman Soto AF} the low-temperature
expansion of the staggered magnetization of O(N) antiferromagnets\cite{Hofmann AF} and
spin waves in canted phases.\cite{Roman Soto Canted} An interesting application
concerning SO(5) invariance and high-$T_c$ superconductivity can be found in
Ref.\cite{Burgess Lutken}. Pedagogical introductions to effective Lagrangians are
Refs.\cite{Leutwyler Brasil and Chiral Dynamics,Burgess,Lectures and Courses}, brief
outlines of the method can be found in Refs.\cite{Brief descriptions}.

As the effective analysis refers to large wavelengths, it does not resolve the
microscopic structure of a solid and the system hence appears homogeneous. Accordingly,
the effective Lagrangian is invariant with respect to translations. On the other hand,
the effective Lagrangian is not invariant under rotations, since the lattice structure
of a solid singles out preferred directions. In the case of a cubic lattice, the
anisotropy, however, only shows up at higher orders of the derivative
expansion:\cite{Hasenfratz Niedermayer 1993,Roman Soto} the discrete symmetries of a
cubic lattice thus imply space rotation symmetry. In the following, we assume that the
ferromagnet exhibits this type of lattice structure: the leading order effective
Lagrangian is then invariant both under translations and under rotations and the
corresponding expression for a ferromagnet takes the form\cite{Leutwyler NRD,Roman Soto}
\begin{equation}
\label{Leffp2}
{\cal L}^2_{eff} \ = \ \Sigma \, \frac{{\varepsilon}_{ab} \, {\partial}_0 U^a
U^b}{1 + U^3} \; + \; {\Sigma} \mu H U^3 \; - \; \mbox{$ \frac{1}{2}$} F^2
{\partial}_rU^i {\partial}_rU^i \, .
\end{equation}
In the above notation, the two real components of the magnon field, $U^a (a = 1,2)$, have
been collected in a three-dimensional unit vector $U^i = (U^a, U^3)$, which transforms
with the vector representation of the rotation group O(3). At leading order, the
ferromagnet is thus characterized by two different low energy constants, $\Sigma$ and
$F$. The first term, which involves a time derivative, is related to a topological
invariant. Remarkably, due to this contribution proportional to the spontaneous
magnetization $\Sigma$, the effective Lagrangian of a ferromagnet fails to be invariant
under the group G = O(3). Note that such expressions, involving order parameters
associated with the generators of the spontaneously broken group, would not be permitted
in Lorentz-invariant effective theories. It represents the main novelty occurring in
condensed matter physics, where nonrelativistic kinematics is less restrictive than
Lorentz invariance.

The associated equation of motion is the Landau-Lifshitz equation,
\begin{equation}
\label{equationOfMotion}
{\partial}_0 U^a \, + \, {\varepsilon}_{ajk} f^j_0 U^k \, + \, \gamma
{\varepsilon}_{ajk} \Delta U^j U^k \; = \; 0 \, , \qquad f^j_0 = \mu H {\delta}^j_3 \, ,
\qquad \gamma \equiv \frac{F^2}{\Sigma} \, ,
\end{equation}
which describes the dynamics of ferromagnetic spin waves. Its structure is of the 
Schr\"odinger type: first order in time, but second order in space. Note that, according
to Goldstone's theorem, there have to be two real Goldstone fields in the case of a
spontaneously broken symmetry O(3) $\to$ O(2). However, in the nonrelativistic regime,
this does not necessarily imply that there also have to exist two independent magnon
states. Indeed, in the present case of a ferromagnet, a complex field is required to
describe one particle: there exists only {\it one} type of spin-wave excitation in a O(3)
ferromagnet, following a quadratic dispersion relation, 
\begin{equation}
\label{disprelF}
\omega(\vec k) \, = \, \gamma{\vec k}^2 + {\cal O}(|{\vec k}|^4) \, . 
\end{equation}
Accordingly, in the effective description of this nonrelativistic system, {\bf two}
powers of momentum correspond to only {\bf one} power of energy or temperature: $k^2
\propto \omega, T$.

The effective Lagrangian method provides us with a simultaneous expansion in powers of
momentum and of the external field $H$. The important point is that, to a given order in
the low-energy expansion, only a finite number of coupling constants and a finite number
of graphs contribute. Let us now set up the effective Lagrangian for a ferromagnet at
higher orders in the momentum expansion\cite{footnote2} and consider the corresponding
power counting scheme.

\section{Power Counting and Effective Lagrangian}
\label{EffectiveLagrangian}

In his pioneering microscopic analysis, Dyson evaluated the temperature expansion for the
partition function of a cubic ferromagnet up to terms of order $T^5$. In our effective
language, where one power of temperature or energy counts like two powers of momentum,
this corresponds to an expansion up to order $p^{10}$. We will now show that, to this
order in the momenta, contributions to the effective Lagrangian involving at most six
space derivatives enter and the perturbative expansion requires the evaluation of graphs
containing at most two loops.

First of all, note that there are no contributions to the effective Lagrangian with an
odd number of space derivatives: parity excludes terms like
\begin{equation} c_{abc} \, {\varepsilon}_{rst} \, {\partial}_r U^a {\partial}_s U^b 
{\partial}_t U^c \, ,
\end{equation}
which involve the antisymmetric tensor ${\varepsilon}_{rst}\;$.

Before constructing the relevant pieces ${\cal L}^{4}_{eff}, {\cal L}^{6}_{eff}, \ldots$
step by step, we have to point out an important assumption underlying the present
analysis: we assume that the effective Lagrangian of a ferromagnet is gauge invariant at
subleading orders. In fact, the discrete symmetries of a cubic lattice and time-reversal
invariance ensure that there is no topological contribution in ${\cal L}^{4}_{eff}$ (see
Ref.\cite{Roman Soto}). Topological contributions which may show up in
${\cal L}^{6}_{eff}$ or beyond do not affect the conclusions of the present work.

Gauge invariance then implies that the magnetic field $H$ enters the derivative expansion
through a timelike covariant derivative,
\begin{equation}
D_0 U^i \; = \; {\partial}_0 U^i \, + \, {\varepsilon}_{ijk} f^j_0 U^k \, ,
\qquad f^j_0 = \mu H {\delta}^j_3 \, .
\end{equation}
Note that the magnetic field occurs on the same level as the time derivative and is thus
counted as a quantity of order $p^2$.

The next-to-leading order Lagrangian is of order $p^4$. It involves terms with two time
derivatives, terms with one time and two space derivatives and terms with four space 
derivatives. Time derivatives, however, can be eliminated with the equation of
motion,\cite{footnote3}
\begin{displaymath}
{\partial}_0 U^a \, + \, {\varepsilon}_{ajk} f^j_0 U^k \, + \, \gamma
{\varepsilon}_{ajk} \Delta U^j U^k \; = \; 0 \, ,
\end{displaymath}
such that we end up with the following independent terms:\cite{footnote4}
\begin{equation}
\label{Leffp4}
{\cal L}^4_{eff} \ = \ l_1 \, ({\partial}_r U^i {\partial}_r U^i)^2 + l_2 \,
({\partial}_r U^i {\partial}_s U^i)^2 + l_3 \, \Delta U^i \Delta U^i \, .
\end{equation}
We thus have three effective coupling constants, $l_1, l_2$ and $l_3$, at next-to-leading
order. Note that all terms involving the magnetic field have been eliminated with the
equation of motion. In particular, gauge invariance implies that there is no tree graph
from ${\cal L}^4_{eff}$ contributing to the vacuum energy.

Clearly, the question now arises, to what order $n$ in the effective Lagrangian
${\cal L}^{2n}_{eff}$ we have to go to evaluate the partition function up to accuracy
$p^{10}$. In the following, we will show that loop corrections involving ferromagnetic
magnons are suppressed by {\it three} powers of momentum and the effective Lagrangian is
needed up to ${\cal L}^6_{eff}$.

To verify this statement, let us consider a scattering process. The effective Lagrangian
provides us with an expansion for a multi-magnon scattering amplitude in powers of
momentum. The elastic magnon-magnon scattering amplitude is obtained by expanding
${\cal L}_{eff}$ up to order $(U^a U^a)^2$. As shown in Figure 1a, the leading term
corresponds to the tree graph involving the four-magnon vertex from ${\cal L}^2_{eff}$
and is of order $F^2{\vec k}^{\,2}/{\Sigma}^2$ (see Ref.\cite{Hofmann spin wave}). The
first correction comes from the tree graph 1b with a vertex involving the next-to-leading
order Lagrangian ${\cal L}^4_{eff}$. However, we also have to consider loop graphs in the
effective framework. These corrections are in general infinite and need to be 
renormalized.

\begin{figure}[!t]
\centerline{\epsfysize=3.5truecm \hbox{\epsfbox{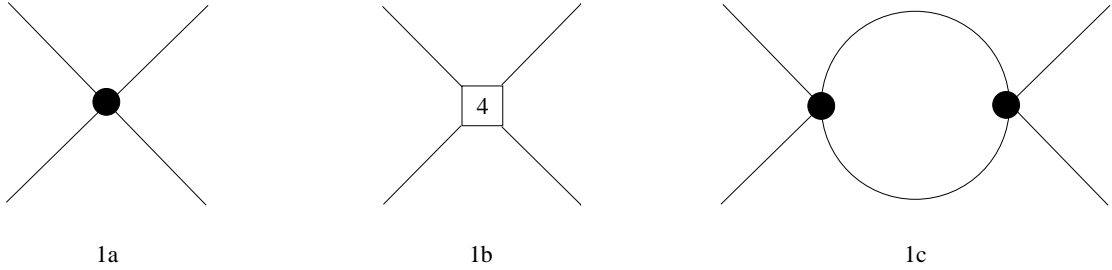}}  }
\medskip
{\tighten \caption[1]{Feynman graphs describing the leading contributions to the
magnon-magnon scattering amplitude. The numbers attached to the vertices refer to the
piece of the effective Lagrangian they come from. Vertices associated with the leading
term ${\cal L}^2_{eff}$ are denoted by a dot.} \label{fd1} }
\end{figure} 

Consider the magnon-magnon scattering graph 1c, where the two vertices come from
${\cal L}^2_{eff}$. The form of the loop graph is
\begin{equation}
\label{CountingRule1}
{\cal A}_{loop} \, \propto \, \int \! \frac{{d} \omega \,
{d}^3\!k}{(2\pi)^4} \, \Bigg(\frac{F^2 {\vec k}^{\,2}}{{\Sigma}^2} \Bigg)^2
\Bigg\{\frac{1}{\omega - \gamma {\vec k}^{\,2}} \Bigg\}^2 \, , 
\end{equation}
where the quantities $\omega$ and ${\vec k}$ denote generic energy and three momentum,
respectively. Each four-magnon vertex from ${\cal L}^2_{eff}$ is of order
$F^2 {\vec k}^{\,2}/{\Sigma}^2$, while the two magnon propagators are of order
$1/(\omega - \gamma {\vec k}^{\,2})$. By dimensional analysis, the loop integral must
then have the form
\begin{equation}
\label{CountingRule2} 
{\cal A}_{loop} \, \propto \, \frac{\gamma}{\Sigma^2} \, {|\vec p \, |}^{\,5} \, .
\end{equation}
Comparing this expression with the leading tree graph amplitude 1a,
\begin{equation}
{\cal A}_{tree} \, \propto \, \frac{\gamma}{\Sigma} \, {\vec p}^{\,2} \, , 
\end{equation}
we find the remarkable result that, in the effective description of the ferromagnet,
loops are suppressed by {\bf three} powers of momentum. This is to be contrasted with the
Lorentz-invariant framework (chiral perturbation theory), where loops are suppressed by
two powers of momentum in four dimensions.

Hence, in order to carry out the expansion for the partition function up to accuracy
$p^{10}$, we need the Lagrangian up to terms of order $p^6$: ${\cal L}^6_{eff}$ shows up
in a one-loop graph at order $p^9$. Notice that the only relevant term coming from 
${\cal L}^6_{eff}$ is quadratic in the magnon field. Eliminating time derivatives, we end
up with 
\begin{eqnarray}
\label{Leffp6-8}
{\cal L}^6_{eff} \ = \ c_1 \, U^i {\Delta}^3 U^i \, .
\end{eqnarray}
As in ${\cal L}^4_{eff}$, terms involving the magnetic field do not show up: they
have all been eliminated with the equation of motion. More generally, assuming
${\cal L}^4_{eff}, {\cal L}^6_{eff}, {\cal L}^8_{eff}$ and ${\cal L}^{10}_{eff}$ to be
gauge invariant, there are no tree-graph contributions from these higher order pieces of
the effective Lagrangian to the vacuum energy -- provided that the magnetic field is the
only external field considered.

\section{Finite Temperature}
\label{Finite T}

In finite temperature field theory, the partition function is represented as a Euclidean
functional integral,\cite{footnote7,Review finite T,Landsman Weert Kapusta Smilga}
\begin{equation}
\label{TempExp}
\mbox{Tr} \, [\exp(- {\cal H}/T)] \, = \, \int [dU] \, \exp \Big(- {\int}_{\!
\! \! {\cal T}} \! \! d^4x \, {\cal L}_{eff}\Big) \, .
\end{equation}
The integration is performed over all field configurations which are periodic in the
Euclidean time direction $U({\vec x}, x_4 + \beta) = U({\vec x}, x_4)$, with
$\beta \equiv 1/T$. The low-temperature expansion of the partition function is obtained
by considering the fluctuations of the field $U$ around the ground state $V = (0, 0, 1)$,
i.e., by expanding $U^3$ in powers of $U^a$, $U^3 = \sqrt{1-U^aU^a}$. The leading
contribution is of order $p^2$ and contains a term  quadratic in $U^a$ which describes
free magnons. In the presence of a magnetic field, they obey the dispersion
relation\cite{footnote8}
\begin{equation}
\label{disprelFH}
i k_4(\vec k) \; = \; \gamma{\vec k}^2 \, + \, \mu H + {\cal O}(|{\vec k}|^4) 
\, .
\end{equation}
The remainder of the effective Lagrangian is treated as a perturbation. Evaluating the
Gaussian integrals in the standard manner, one arrives at a set of Feynman rules which
differ from the conventional rules of the effective Lagrangian method only in one
respect: the periodicity condition imposed on the Goldstone field modifies the
propagator. At finite temperature, the propagator is given by
\begin{equation}
\label{ThermalPropagator}
G(x) \, = \, \sum_{n \,= \, - \infty}^{\infty} \Delta({\vec x}, x_4 + n \beta)
\, ,
\end{equation}
where $\Delta(x)$ is the Euclidean propagator at zero temperature,
\begin{equation}
\label{Propagator}
\Delta (x) \, = \, \int \! \, \frac{d k_4 d^3\!k}{(2\pi)^4}
\frac{e^{i{\vec k}{\vec x} - i k_4 x_4}}{\gamma {\vec k}^2 - i k_4 + \mu H} \, . 
\end{equation}
Note that the above Green function corresponds to the propagation of a single magnon, 
described by the complex field $u = U^1 + i U^2$.

We restrict ourselves to the infinite volume limit and evaluate the free energy 
density $z$, defined by
\begin{equation}
\label{freeEnergyDensity}
z \ = \ -T \, \lim_{L \to \infty} L^{-3} \, \ln \, [ \mbox{Tr}
\exp(- {\cal H}/T)] \, .
\end{equation}
Temperature thus produces remarkably little change: to obtain the partition function, one
simply restricts the manifold on which the fields are living to a torus $\cal T$ in
Euclidean space. The effective Lagrangian remains unaffected -- the coupling constants
$F, \Sigma, l_1, \dots$ are temperature independent.

It is convenient to use dimensional regularization to evaluate the graphs of the
effective theory. Unlike in a Lorentz-invariant framework, where the physical dimension
$d$ is equal to four, we regularize the nonrelativistic propagator only in the three
spatial components,
\begin{equation}
\label{PropagatorDimReg}
\Delta (x) \ = \ \frac{1}{(2{\pi})^d} \, \Big(\frac{{\pi}}{\gamma}\Big)^{\frac{d}{2}}
\frac{1}{{x_4}^{\frac{d}{2}}} \exp{\Big(- \frac{{\vec x}^2}{4 \gamma x_4} \, -
\, \mu H x_4 \Big)} \, \Theta (x_4) \, ,
\end{equation}
and set $d=3$ at the end of the calculation.

For the thermal propagator, we then have
\begin{equation}
\label{ThermProp}
G(x) \ = \ \frac{1}{(2{\pi})^d} \, \Big(\frac{{\pi}}{\gamma}\Big)^{\frac{d}{2}}
\sum^{\infty}_{n \, = \, - \infty} \frac{1}{x_n^{\frac{d}{2}}} \, \exp{\Big(- \frac{{\vec
x}^2}{4 \gamma x_n} \, - \, \mu H x_n \Big)} \, \Theta (x_n) \, ,
\end{equation}
with
\begin{equation}
x_n \, \equiv \, x_4 + n \beta \, .
\end{equation}
Using complex notation for the magnon field, the evaluation of Gaussian integrals
simplifies, since the following two-point functions vanish:
\begin{equation}
\label{PathIntZero}
\langle 0 | \, T\{u({\vec x}, x_4) \, u({\vec y}, y_4)\} | 0 \rangle \, =
\, \langle 0 | \, T\{u^{\dagger}({\vec x}, x_4) \, u^{\dagger}({\vec y}, y_4)\} | 0
\rangle\ \, = \, 0 \, . 
\end{equation}
Nonzero contributions to the partition function only result from
\begin{equation}
\label{PathInt1}
\langle 0 | \, T\{u({\vec x}, x_4) \, u^{\dagger}({\vec y}, y_4)\} | 0 \rangle \, = \,
\frac{2}{\Sigma} \, \Delta (x-y) \, .
\end{equation}
The calculation simplifies further due to space rotation symmetry of the propagator,
implying that, at the origin, single space derivatives vanish,
\begin{equation}
\label{behaviorProp2}
\qquad {\partial}_r G(|{\vec x}|, x_4)_{|x =0} \; = \; 0 \, .
\end{equation}
Let us now evaluate the free energy density order by order in the momentum expansion.

\section{Evaluation of the Free Energy Density}
\label{Partition Function}

The relevant Feynman graphs are shown in Figure 2. Depicted are all contributions to the
free energy density up to order $p^{10}$ or, equivalently, order $T^5$.

\begin{figure}[!t]
\centerline{\epsfysize=6.0truecm \hbox{\epsfbox{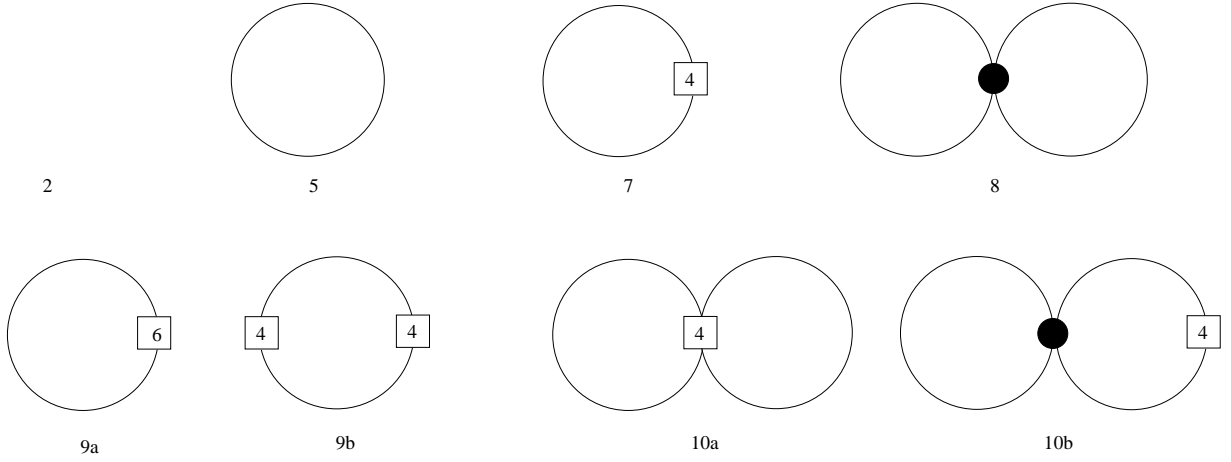}}  }
\medskip
{\tighten \caption[2]{Feynman graphs occurring in the low-temperature expansion of
the partition function for an O(3) ferromagnet up to order $p^{10}$. The numbers attached
to the vertices refer to the piece of the effective Lagrangian they come from. Vertices
associated with the leading term ${\cal L}^2_{eff}$ are denoted by a dot. The numbers
specifying individual graphs correspond to the power of momentum in the derivative
expansion. Note that, in the counting scheme for ferromagnetic magnons, two powers of
momentum correspond to one power of energy or temperature.} \label{fd2} }
\end{figure} 

As we will see, all these contributions to the free energy density exclusively involve
propagators, and space derivatives thereof, to be evaluated at the {\it origin}
$x\!=\!0$. It is convenient to introduce the following notation:
\begin{eqnarray}
\label{definitionsThermProp}
G_1 \ & \equiv & \ \, \Big[ G(x) \Big]_{x=0} \, , \nonumber \\
G_{\Delta} & \equiv & \, \Big[ {\Delta} G(x) \Big]_{x=0} \, ,
\end{eqnarray}
where $\Delta$ represents the Laplace operator in three dimensions -- no confusion should
occur with $\Delta (x)$, denoting the zero-temperature propagator.

The quantity $G_1$ is split into a finite piece, which is temperature dependent, and a
divergent piece, which is temperature independent,
\begin{equation}
\label{DecompositionPropagator}
G_1 \; = \; G^T_1 \, + \, G^0_1 \, .
\end{equation}
Using dimensional regularization, the explicit expressions are
\begin{eqnarray}
\label{ThermProp(x=0)}
G^T_1 & = & \frac{1}{(2{\pi})^d} \,
\Big(\frac{\pi}{\gamma}\Big)^{\frac{d}{2}} \, \sum^{\infty}_{n=1}
\frac{e^{- \mu H n \beta}}{(n \beta)^{\frac{d}{2}}} \, , \nonumber \\
G^0_1 & = & \frac{1}{(2{\pi})^d} \, \Big(\frac{{\pi}}{\gamma}\Big)^{\frac{d}{2}}
\Bigg [\frac{1}{{x_4}^{\frac{d}{2}}} \exp\Big(-\frac{{\vec x}^2}{4 \gamma x_4} \Big)
\Theta(x_4) \Bigg]_{x=0} \, .
\end{eqnarray}
Likewise, for $G_{\Delta}$ we have
\begin{eqnarray}
\label{DeltaProp(x=0)}
G^T_{\Delta} & = & \frac{1}{(2{\pi})^d} \, \Big(\frac{\pi}{\gamma}\Big)^{\frac{d}{2}} \,
\Big(\! - \frac{d}{2\gamma} \Big) \sum^{\infty}_{n=1} \frac{e^{ - \, \mu H n
\beta}}{(n\beta)^{\frac{d}{2} + 1}} \, , \nonumber \\
G^0_{\Delta} & = & \frac{1}{(2{\pi})^d} \, \Big(\frac{\pi}{\gamma}\Big)^{\frac{d}{2}}
\Bigg[ \frac{1}{{x_4}^{\frac{d}{2} + 1}} \Bigg\{ \frac{-d}{2 \gamma} + \frac{{\vec x}^2}
{4 {\gamma}^2 x_4} \Bigg\} \exp\Big(-\frac{{\vec x}^2}{4 \gamma x_4} \Big)
\Theta(x_4) \Bigg]_{x=0} \, .
\end{eqnarray}
The temperature-independent pieces $G^0_1$ and $G^0_{\Delta}$, as well as propagators
involving higher orders of space derivatives, are all related to momentum integrals of
the form
\begin{equation}
\label{RuleDimReg}
\int \! \, d^d\!k \, \Big( {\vec k}^2 \Big)^m \exp\Big[ - \gamma x_4 {\vec k}^2 - x_4
\mu H \Big] \, , \qquad m = 0, 1, 2, \ldots \, ,
\end{equation}
which are proportional to
\begin{equation}
\label{RuleDimReg2}
\frac{\exp[- x_4 \mu H]}{(\gamma x_4)^{m+\frac{d}{2}}} \, \, \Gamma(m+\frac{d}{2}) \, .
\end{equation}
In dimensional regularization, however, these expressions vanish altogether: $G^0_1$,
$G^0_{\Delta}$ and propagators involving more Laplacians do not contribute in the limit
$d\!\to\!3$. Therefore, in the evaluation of the above graphs, only the
temperature-dependent pieces $G^T_1, G^T_{\Delta}, \dots$ are relevant.

The first contribution to the free energy density, which is of order $p^2$, comes from
tree graph 2. It does not depend on temperature and is given by
\begin{equation}
\label{z(2)}
z_{2} \ = \ - \Sigma \mu H \, .
\end{equation}

The order $p^5$ contribution from one-loop graph 5 is associated with a $d$-dimensional
nonrelativistic free Bose gas. For the  temperature-dependent part, we obtain
\begin{equation}
\label{z(5)T}
z^T_{5} \ = \ - \frac{1}{8 {\pi}^{\frac{3}{2}} {\gamma}^{\frac{3}{2}}}
\, T^{\frac{5}{2}} \sum^{\infty}_{n=1} \frac{e^{- \mu H n \beta}}{n^{\frac{5}{2}}} \, .
\end{equation}

At order $p^7$, the next-to-leading order Lagrangian ${\cal L}^4_{eff}$ comes into play.
For one-loop graph 7, which involves a two-magnon vertex, we get 
\begin{equation}
\label{z(7)}
z_{7} \ = \ - \frac{2 \, l_3}{{\Sigma}} \, \Big[ {\Delta}^2 G(x) \Big]_{x=0} \, ,
\end{equation}
yielding
\begin{equation}
\label{z(7)T}
z^T_{7} \ = \ - \frac{15 \, l_3}{16 {\pi}^{\frac{3}{2}} \Sigma {\gamma}^{\frac{7}{2}}} \,
T^{\frac{7}{2}} \sum^{\infty}_{n=1} \frac{e^{- \mu H n \beta}}{n^{\frac{7}{2}}} \, \, .
\end{equation}

The first two-loop graph appears at order $p^8$. Remarkably, this contribution is
proportional to single space derivatives of the propagator at the origin and thus
vanishes,
\begin{equation}
\label{z(8)}
z_{8} \ \propto \, \Big[ {\partial}_r G(x) \Big]_{x=0} \,
\Big[{\partial}_r G(x) \Big]_{x=0} \ = \ 0 \, .
\end{equation}

At order $p^9$, two one-loop graphs show up which involve ${\cal L}^4_{eff}$ and ${\cal 
L}^6_{eff}$, respectively. For graph 9a, we get
\begin{equation}
\label{z(9a)}
z_{9a} \ = \ - \frac{2 \, c_1}{\Sigma} \, \Big[ {\Delta}^3 G(x) \Big]_{x=0} \, ,
\end{equation}
leading to the temperature-dependent contribution
\begin{equation}
\label{z(9a)T}
z^T_{9a} \ = \ \frac{105 \, c_1}{32 {\pi}^{\frac{3}{2}} \Sigma {\gamma}^{\frac{9}{2}}} \,
T^{\frac{9}{2}} \sum^{\infty}_{n=1} \frac{e^{- \mu H n \beta}}{n^{\frac{9}{2}}} \, \, .
\end{equation}
Graph 9b is proportional to an integral over the torus $\cal T$ which involves a product
of two propagators,
\begin{equation}
\label{z(9b)}
z_{9b} \, = \, - \frac{2 \, l^2_3}{{\Sigma}^2} \, {\int}_{\! \! \! {\cal T}} \! \!
d^{d+1}x \, {\Delta}^2 G(x) \, {\Delta}^2 G(-x) \, .
\end{equation}
This expression, however, can be reduced to an $x$-independent term involving one
propagator only. To verify this statement, consider the equation for the thermal
propagator,
\begin{equation}
\label{PropEquation}
\Bigg( \frac{{\partial}}{{\partial} x_4} - \gamma \Delta + \mu H \Bigg) \, G(x)
\, = \, \delta(x) \, ,
\end{equation}
and take the derivative with respect to the magnetic field. The quantity 
$\partial G(x) / \partial (\mu H)$ may then be written as a convolution integral over the
torus. At the origin, the expression reads
\begin{equation}
\label{2->1}
\Bigg[ \frac{\partial G(x)}{\partial (\mu H)} \Bigg]_{x=0} \, = \, - {\int}_{\! 
\! \! {\cal T}} \! \! d^{d+1}y \, \, G(-y) \, G(y) \, .
\end{equation}
Inserting Laplace operators at intermediate steps, we obtain the more general result
\begin{equation}
\label{2->1Delta4}
\Bigg[ {\Delta}^{(m+n)} \frac{\partial G(x)}{\partial (\mu H)} \Bigg]_{x=0}
\, = \, - {\int}_{\! \! \! {\cal T}} \! \! d^{d+1}y \, {\Delta}^m G(-y) \, {\Delta}^n
G(y) \, ,
\end{equation}
such that (\ref{z(9b)}) can be written as
\begin{equation}
z_{9b} \, = \, \frac{2 \, l^2_3}{{\Sigma}^2} \, \Bigg[ {\Delta}^4 
\frac{\partial G(x)}{\partial (\mu H)} \Bigg]_{x=0} \, .
\end{equation}
For the temperature-dependent part of graph 9b, we then get
\begin{equation}
\label{z(9b)T}
z^T_{9b} \ = \ - \frac{945 \, l_3^2}{64 {\pi}^{\frac{3}{2}} {\Sigma}^2
{\gamma}^{\frac{11}{2}}} \, T^{\frac{9}{2}} \sum^{\infty}_{n=1}
\frac{e^{- \mu H n \beta}}{n^{\frac{9}{2}}} \, \, .
\end{equation}

Finally, at order $p^{10}$, two-loop graphs with insertions from ${\cal L}^4_{eff}$ show 
up. Graph 10a contributes with
\begin{equation}
\label{z(10a)}
z_{10a} = - \frac{2}{3 {\Sigma}^2} (8 l_1 + 6 l_2 + 5 l_3)
\, G_{\Delta} G_{\Delta} \, - \frac{2 \, l_3}{{\Sigma}^2} \, G_1 \Big[ {\Delta}^2 
G(-x) \Big]_{x=0} \, .
\end{equation}
The evaluation of graph 10b amounts to
\begin{equation}
\label{z(10b)}
z_{10b} = \frac{2 \, l_3}{{\Sigma}^2} G_1 \Big[ {\Delta}^2 G(-x) \Big]_{x=0} \, ,
\end{equation}
and cancels the second term in $z_{10a}$. For the temperature-dependent part of order
$p^{10}$ we thus end up with
\begin{equation}
\label{z(10)}
z_{10} = - \frac{3 (8 l_1 + 6 l_2 + 5 l_3)}{128 {\pi}^3
{\Sigma}^2 {\gamma}^5} \, T^5 {\Bigg\{ \sum^{\infty}_{n=1} \frac{e^{- \mu H n
\beta}}{n^{\frac{5}{2}}} \Bigg\}}^2 \, \, .
\end{equation}

Collecting terms, the result for the free energy density up to order $p^{10}$ becomes
\begin{eqnarray}
\label{FreeCollect}
z \; = \; - \Sigma \mu H \; - \; \frac{1}{8 {\pi}^{\frac{3}{2}} {\gamma}^{\frac{3}{2}}} \
T^{\frac{5}{2}} \sum^{\infty}_{n=1} \frac{e^{- \mu H n \beta}}{n^{\frac{5}{2}}}
\; - \; \frac{15 \,l_3}{16 {\pi}^{\frac{3}{2}} \Sigma {\gamma}^{\frac{7}{2}}} \
T^{\frac{7}{2}} \sum^{\infty}_{n=1} \frac{e^{- \mu H n \beta}}{n^{\frac{7}{2}}}
\hspace{1.6cm} \nonumber \\
- \; \frac{105}{32 {\pi}^{\frac{3}{2}} \Sigma {\gamma}^{\frac{9}{2}}} \,
\Bigg( \frac{9 l^2_3}{2 \gamma \Sigma} - c_1 \Bigg) \ T^{\frac{9}{2}}
\sum^{\infty}_{n=1} \frac{e^{- \mu H n \beta}}{n^{\frac{9}{2}}} \hspace{3cm} \nonumber \\
- \; \frac{3 (8 l_1 + 6 l_2 + 5 l_3)}{128 {\pi}^3 {\Sigma}^2 {\gamma}^5} \ T^5
{\Bigg\{ \sum^{\infty}_{n=1} \frac{e^{- \mu H n\beta}}{n^{\frac{5}{2}}} \Bigg\}}^2
\, . \hspace{3cm}
\end{eqnarray}
The first term is temperature-independent and originates from tree graph 2. Contributions
which involve half integer powers of the temperature -- $T^{5/2}, T^{7/2}$ and
$T^{{9}/2}$, respectively -- arise from one-loop graphs and are all related to the free
energy density of noninteracting magnons. Remarkably, there is only one term in the above
series, the order $T^5$ contribution coming from two-loop graph 10a, which is due to the
magnon-magnon interaction.

\section{Interlude: Free Magnons versus Interacting Magnons}
\label{Interaction}

Consider the formula for the free energy density of a gas of noninteracting bosons,
\begin{equation}
\label{FreeFunction2}
z \; = \; z_0 \, + \, \frac{T}{(2 \pi)^3} \, \int \!\! d^3 \! k \, \ln
\Big( 1 - e^{- \omega({\vec k}) / T} \Big) \, ,
\end{equation}
where $z_0$ is the energy density of the vacuum. With the leading term of the dispersion
relation,
\begin{displaymath}
\omega(\vec k) \, = \, \gamma{\vec k}^2 + \mu H + {\cal O}(|{\vec k}|^4) \, , \qquad
\quad \gamma \equiv \frac{F^2}{\Sigma} \, ,
\end{displaymath}
we readily reproduce the dominant one-loop contribution $z^T_5 \propto T^{5/2}$ in the
free energy density of the O(3) ferromagnet.

\begin{figure}[!t]
\centerline{\epsfysize=9.0truecm \hbox{\epsfbox{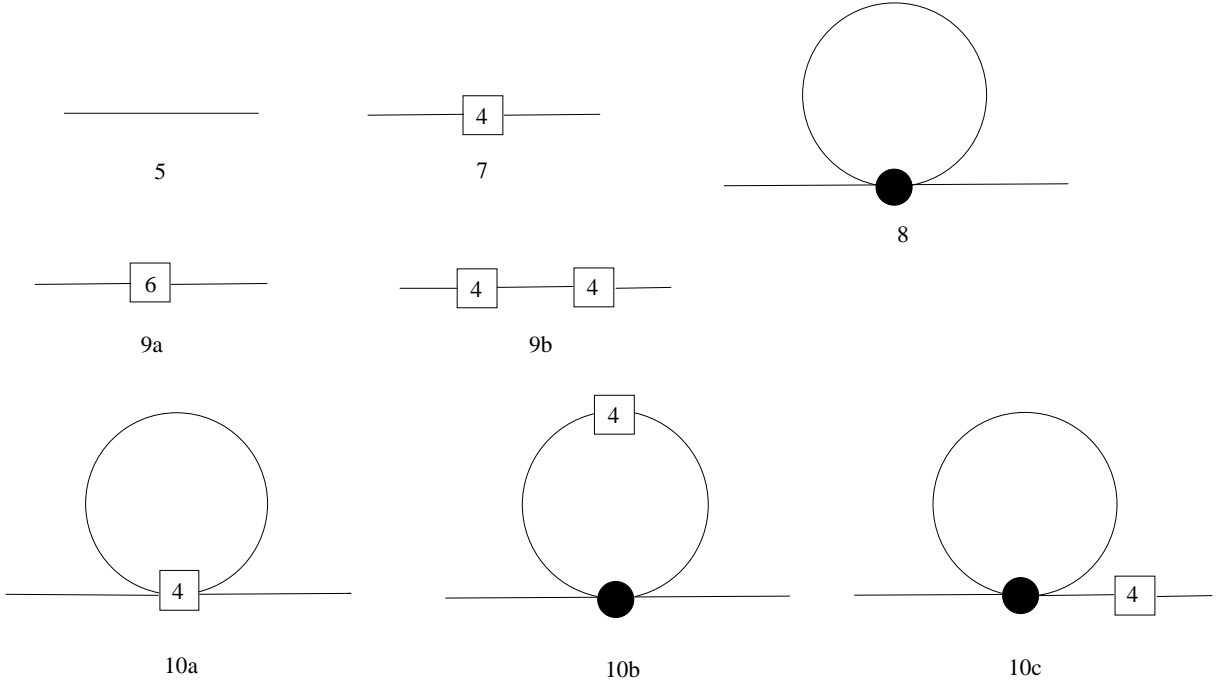}}  }
\medskip
{\tighten \caption[3]{Feynman graphs occurring in the low-energy expansion of the
two-point function for an O(3) ferromagnet up to order $p^{10}$. The numbers specifying
individual graphs correspond to the power of momentum in the derivative expansion.}
\label{fd3} }
\end{figure}

In order to account for subleading terms in the dispersion relation, we have to evaluate
higher order contributions to the two-point function $\langle 0 | \, T\{u({\vec x}, x_4)
\, u^{\dagger}({\vec y}, y_4)\} | 0 \rangle$ of the magnon field. The relevant graphs are
shown in Figure 3. Depicted are all contributions up to order $p^{10}$. Instead of
listing individual results for the two-point function, we give the final expression for
the dispersion relation originating from these graphs:
\begin{equation}
\label{DispRel}
\omega(\vec k) \; = \; \frac{1}{\Sigma} \Big( F^2 {\vec k}^2 \, + \, \Sigma \mu H \,
- \, 2 l_3 {\vec k}^4 \, + \, 2 c_1 {\vec k}^6 \Big) \, .
\end{equation}
The two new terms correspond to graph 7, graph 9a and graph 9b of Figure 3. Note that
one-loop graph 8 does not contribute to the dispersion relation. The effect of the three
remaining one-loop graphs of order $p^{10}$ will be discussed in a moment.

Inserting the above dispersion relation into formula (\ref{FreeFunction2}), we reproduce
all contributions to the free energy density which are associated with one-loop graphs:
graphs 5, 7, 9a and 9b of Figure 2. These contributions are all related to noninteracting
magnons -- the corresponding coefficients in the dispersion relation are independent of
the temperature.

As far as the evaluation of the three one-loop graphs of order $p^{10}$ (Figure 3) is
concerned, the situation is the following: in the sum of these contributions, only one
term remains. It originates from graph 10a and leads to an additional term in the
dispersion relation
\begin{equation}
\label{DispRelRen}
\propto \frac{\Big( 8 l_1 + 6 l_2 + 5 l_3 \Big)}{{\Sigma}^2} \, G^0_{\Delta} \;
{\vec k}^2 \, .
\end{equation}
However, in dimensional regularization, the quantity $G^0_{\Delta}$, as we have seen,
vanishes, such that the dispersion relation is not affected by the graphs of order
$p^{10}$.

Nevertheless, we can mimic the magnon-magnon interaction by allowing some of the
coefficients of the dispersion relation to become temperature dependent. If we replace
$G^0_{\Delta}$ by $G^T_{\Delta}$ in (\ref{DispRelRen}), we obtain an interaction term
proportional to $T^5$ in the free energy density. Moreover, we see that the leading order
coupling constant $\gamma$ is not renormalized at leading order ($\propto G^T_1
\,{\vec k}^2$, i.e., order $p^8$ in the free energy density), but only at next-to-leading
order ($\propto G^T_{\Delta} \,{\vec k}^2$, i.e., order $p^{10}$ in the free energy
density). This observation, implying that the two-loop graph 8 does not contribute to the
dispersion relation, was already made in Refs.\cite{Pisarski}.

In this section, we have split the free energy density into a piece corresponding to
free magnons and a piece representing the magnon-magnon interaction,
\begin{equation}
\label{zFreeInt}
z \; = \; z_{free} \, + \, z_{inter} \, .
\end{equation}
The free part can be obtained from the free Bose gas formula
\begin{displaymath}
z_{free} \; = \; z_0 \, + \, \frac{T}{(2 \pi)^3} \, \int \!\! d^3 \! k \, \ln
\Big( 1 - e^{- \omega({\vec k}) / T} \Big) \, ,
\end{displaymath}
with the modified dispersion relation
\begin{displaymath}
\omega({\vec k}) \; = \; \frac{1}{\Sigma} \Bigg\{ F^2 {\vec k}^2 \, + \, \Sigma \mu H \,
- \, 2 l_3 {\vec k}^4 \, + \, 2 c_1 {\vec k}^6 \Bigg\} \, .
\end{displaymath}
The coupling constants $F, \Sigma, l_3$ and $c_1$ are all independent of the temperature.
The interaction part originates from two-loop graph 10a which involves a four-magnon
vertex from ${\cal L}^4_{eff}$,
\begin{displaymath}
z_{inter} \; = \; - \; \frac{3 (8 l_1 + 6 l_2 + 5 l_3)}{128 {\pi}^3
{\Sigma}^2 {\gamma}^5} \, {\Bigg( \sum^{\infty}_{n=1} \frac{e^{- \mu H n
\beta}}{n^{\frac{5}{2}}} \Bigg)}^2 \, T^5 \, .
\end{displaymath}

In order to make the structure of the low-temperature expansion for the free energy
density of an O(3) ferromagnet more transparent, we rewrite the series in the form
\begin{equation}
\label{FreeCollectStructure}
z \; = \; z_0 \, - \, h_0 \, T^{\frac{5}{2}} \, - \, h_1 \, T^{\frac{7}{2}} \,
- \, h_2 \, T^{\frac{9}{2}} \, - \, h_3 \, T^5 + {\cal O}(T^{\frac{11}{2}}) \, ,
\end{equation}
where the temperature-dependent coefficients $h_i$ are given by
\begin{eqnarray}
\label{FreeCollectT}
z_0 & = & - \Sigma \mu H\, , \nonumber \\
h_0 & = & \frac{1}{8 {\pi}^{\frac{3}{2}} {\gamma}^{\frac{3}{2}}} \, \sum^{\infty}_{n=1}
\frac{e^{- \mu H n \beta}}{n^{\frac{5}{2}}} \, , \nonumber \\
h_1 & = & \frac{15 \, l_3}{16 {\pi}^{\frac{3}{2}} \Sigma {\gamma}^{\frac{7}{2}}} \,
\sum^{\infty}_{n=1} \frac{e^{- \mu H n \beta}}{n^{\frac{7}{2}}} \, , \nonumber \\
h_2 & = & \frac{105}{32 {\pi}^{\frac{3}{2}} \Sigma {\gamma}^{\frac{9}{2}}} \, \Bigg(
\frac{9 l^2_3}{2 \gamma \Sigma} \, - \, c_1 \Bigg) \, \sum^{\infty}_{n=1}
\frac{e^{- \mu H n \beta}}{n^{\frac{9}{2}}} \, , \nonumber \\
h_3 & = & \; \frac{3 (8 l_1 + 6 l_2 + 5 l_3)}{128 {\pi}^3
{\Sigma}^2 {\gamma}^5} \, {\Bigg(
\sum^{\infty}_{n=1} \frac{e^{- \mu H n \beta}}{n^{\frac{5}{2}}} \Bigg)}^2 \, .
\end{eqnarray}
In the limit $(\mu H / T) \! \to \! 0$, these coefficients become temperature independent
and the sums reduce to Riemann zeta functions,
\begin{eqnarray}
\label{FreeCollectT(H=0)}
{\tilde h_0} & = & \frac{1}{8 {\pi}^{\frac{3}{2}} {\gamma}^{\frac{3}{2}}} \,
\zeta(\mbox{$ \frac{5}{2}$})
\, , \nonumber \\
{\tilde h_1} & = & \frac{15 \, l_3}{16 {\pi}^{\frac{3}{2}} \Sigma {\gamma}^{\frac{7}{2}}}
\, \zeta(\mbox{$ \frac{7}{2}$}) \, , \nonumber\\
{\tilde h_2} & = & \frac{105}{32 {\pi}^{\frac{3}{2}} \Sigma {\gamma}^{\frac{9}{2}}} \,
\Bigg( \frac{9 l^2_3}{2 \gamma \Sigma} \, - \, c_1 \Bigg) \,
\zeta(\mbox{$ \frac{9}{2}$}) \, , \nonumber \\
{\tilde h_3} & = & \; \frac{3 (8 l_1 + 6 l_2 + 5 l_3)}{128 {\pi}^3
{\Sigma}^2 {\gamma}^5} \, \zeta^2(\mbox{$ \frac{5}{2}$}) \, .
\end{eqnarray}
 
\section{Spontaneous Magnetization}
\label{SpontMagnet}

With the above representation for the free energy density, the low-temperature series for
some relevant thermodynamic quantities can readily be derived. Since the system is
homogeneous, the pressure is given by the temperature dependent part of the free energy
density,
\begin{equation}
\label{Pz}
P \, = \, z_0 - z \, ,
\end{equation}
and the low-temperature expansion amounts to
\begin{equation}
\label{Pressure}
P \; = \; h_0 \, T^{\frac{5}{2}} \, + \, h_1 \, T^{\frac{7}{2}} \, + \, h_2 \,
T^{\frac{9}{2}} \,
+ \, h_3 \, T^5 + {\cal O}(T^{\frac{11}{2}}) \, ,
\end{equation}
with coefficients $h_i$ given in (\ref{FreeCollectT}).

The first contribution represents the free Bose gas term which originates from one-loop
graph 5. Subsequent terms exhibiting half integer powers of the temperature are related
to the shape of the dispersion curve and represent effects due to the discreteness of the
cubic lattice. Remarkably, there is no $T^4$ term in this series: two-loop graph 8 does
not contribute. In the low-temperature expansion for the pressure, the interaction among
ferromagnetic spin waves only starts manifesting itself at order $T^5$.

The corresponding expressions for the energy density $u$, for the entropy density $s$ and
for the heat capacity $c_V$ are readily worked out from the thermodynamic relations
\begin{equation}
\label{Thermodynamics}
s \, = \, \frac{{\partial}P}{{\partial}T} \, , \qquad u \, = \, Ts - P \, ,
\qquad c_V \, = \, \frac{{\partial}u}{{\partial}T} \, = \, T \,
\frac{{\partial}s}{{\partial}T} \, .
\end{equation}
Reordering powers of the temperature and taking the limit $(\mu H / T) \! \to \! 0$, we
get
\begin{eqnarray}
\label{ThermodynQuantities}
u & = & \mbox{$ \frac{3}{2}$} \, {\tilde h_0} \, T^{\frac{5}{2}} \,
+ \, \mbox{$ \frac{5}{2}$} \, {\tilde h_1} \,  T^{\frac{7}{2}} \,
+ \, \mbox{$ \frac{7}{2}$} \, {\tilde h_2} \, T^{\frac{9}{2}} \,
+ \, 4 {\tilde h_3} \, T^5 + {\cal O}(T^{\frac{11}{2}})
\, , \nonumber \\
s & = & \mbox{$ \frac{5}{2}$} \, {\tilde h_0} \, T^{\frac{3}{2}} \,
+ \, \mbox{$ \frac{7}{2}$} \, {\tilde h_1} \, T^{\frac{5}{2}} \,
+ \, \mbox{$ \frac{9}{2}$} \, {\tilde h_2} \, T^{\frac{7}{2}} \,
+ \, 5 {\tilde h_3} \, T^4 + {\cal O}(T^{\frac{9}{2}})
\, , \nonumber \\
c_V & = & \mbox{$ \frac{15}{4}$} \, {\tilde h_0} \, T^{\frac{3}{2}}
\, + \, \mbox{$ \frac{35}{4}$} \, {\tilde h_1} \, T^{\frac{5}{2}}
\, + \, \mbox{$ \frac{63}{4}$} \, {\tilde h_2} \, T^{\frac{7}{2}} \,
+ \, 20 {\tilde h_3} \, T^4 + {\cal O}(T^{\frac{9}{2}}) \, ,
\end{eqnarray}
with coefficients $\tilde h_i$ given in (\ref{FreeCollectT(H=0)}).

Let us now consider the low-temperature expansion for the spontaneous magnetization,
\begin{equation}
\label{Magnetization}
\Sigma(T) = - \lim_{H\!\to0} \, \frac{\partial z}{\partial (\mu H)} \, .
\end{equation}
With the expression (\ref{FreeCollectStructure}) for the free energy density, we obtain the
following series:
\begin{equation}
\label{MagnetizationTempExtended}
\Sigma(T) \, / \, \Sigma \ = \ 1 \, - \, {\alpha}_0 T^{\frac{3}{2}} \, - \, {\alpha}_1
T^{\frac{5}{2}} \, - \, {\alpha}_2 T^{\frac{7}{2}} \, - \, {\alpha}_3 T^4 \,
+ \, {\cal O} (T^{{\frac{9}{2}}}) \, .
\end{equation}
The coefficients $\alpha_n$ are independent of the temperature and given by
\begin{eqnarray}
\label{SigmaCollectT(H=0)}
{\alpha}_0 & = & \frac{1}{8 {\pi}^{\frac{3}{2}} \Sigma {\gamma}^{\frac{3}{2}}} \,
\zeta(\mbox{$ \frac{3}{2}$}) \, , \nonumber \\
{\alpha}_1 & = & \frac{15 \, l_3}{16 {\pi}^{\frac{3}{2}} {\Sigma}^2
{\gamma}^{\frac{7}{2}}} \, \zeta(\mbox{$ \frac{5}{2}$}) \, , \nonumber \\
{\alpha}_2 & = & \frac{105}{32 {\pi}^{\frac{3}{2}} {\Sigma}^2 {\gamma}^{\frac{9}{2}}} \,
\Bigg( \frac{9 l^2_3}{2 \gamma \Sigma} \, - \, c_1 \Bigg) \,
\zeta(\mbox{$ \frac{7}{2}$}) \, , \nonumber\\
{\alpha}_3 & = & \frac{3 (8 l_1 + 6 l_2 + 5 l_3)}{64 {\pi}^3 {\Sigma}^3 {\gamma}^5}\
\zeta(\mbox{$ \frac{5}{2}$}) \, \zeta(\mbox{$ \frac{3}{2}$}) \, .
\end{eqnarray}
Here comes the appropriate place where we would like to compare our result with Dyson's
microscopic calculation. In fact, one of the main motivations for his analysis was the
following question: at what order in the low-temperature expansion does the spin-wave
interaction manifest itself in the spontaneous magnetization? Before his rigorous
analysis, in which he showed that the spin-wave interaction only starts contributing at
order $T^4$, there appeared to be an amazing mess in the literature. At least three
different answers were available, all of them in contradiction with each other and, as it
turned out, also in contradiction with Dyson's result: the
authors\cite{Kramers Opechowski} obtained a term proportional to $T^2$ and the two
references\cite{Schafroth,Kranendonk} ended up with two different contributions of order
$T^{7/4}$.

Within the effective Lagrangian framework we clearly confirm Dyson's finding: in the
spontaneous magnetization, the leading contribution coming from the magnon-magnon
interaction is of order $T^4$. Any microscopic information, such as the specification of
the type of the cubic lattice, is comprised in the numerical values of the effective
coefficients ${\alpha}_n$ -- they involve the coupling constants $F, \Sigma, l_1, \ldots$
occurring in the derivative expansion of the effective Lagrangian ${\cal L}^2_{eff},
{\cal L}^4_{eff}$ and ${\cal L}^6_{eff}$, which phenomenologically parametrize the
microscopic detail.

After Dyson's work, various authors tried to derive his results with alternative methods
within the microscopic framework of the Heisenberg model. In particular, there was
emerging interest in simplifying his complicated calculation.\cite{Rederivation} In the
course of these investigations, however, not all authors could confirm Dyson's findings
and, surprisingly, a new term in the temperature expansion for the spontaneous
magnetization started to haunt the literature: various authors\cite{Wrong Coefficient}
ended up with an interaction term of order $T^3$. However, the $T^3$ contribution turned
out to be a spurious effect.\cite{Final Agreement} It emerged due to the approximation
methods used in these involved microscopic calculations: random phase approximation and
decoupling approximation in the method of double-time temperature-dependent Green
functions.

We would like to emphasize that there are no such approximations in the effective
Lagrangian framework. The method applies to any system where the Goldstone bosons are the
only excitations without energy gap. In the effective Lagrangian perspective, the problem
is approached from a unified and model-independent point of view, based on the symmetries
inherent in the underlying theory. In the present case of a ferromagnet, it is the
spontaneously broken internal symmetry O(3) $\to$ O(2) of the Heisenberg model and the
nonzero value for the spontaneous magnetization, which dictate the structure of the
low-temperature series considered in this paper.

Our calculation clearly shows that there is no $T^3$ term in the low-temperature
expansion for the spontaneous magnetization of a cubic O(3) ferromagnet. Using effective
language this translates into the statement that the contribution from two-loop graph 8
of Figure 2 to the free energy density vanishes. As we have pointed out, the reason is
due to space rotation symmetry of the leading order effective Lagrangian, implying that
single space derivatives of the propagator, evaluated at the origin, are zero.

The argument is not restricted to cubic lattices. Consider the leading order Lagrangian
of the O(3) ferromagnet:
\begin{displaymath}
{\cal L}^2_{eff} \ = \ \Sigma \, \frac{{\varepsilon}_{ab} \, {\partial}_0 U^a
U^b}{1 + U^3} \; + \; {\Sigma} \mu H U^3 \; - \; \mbox{$ \frac{1}{2}$} F^2
{\partial}_rU^i {\partial}_rU^i \, .
\end{displaymath}
Let us replace the last term, which is invariant under space rotations, by the more
general expression
\begin{equation}
\label{AnisotropicLagrangian}
- \; \mbox{$ \frac{1}{2}$} F_1^2 {\partial}_1U^i {\partial}_1U^i \;
- \; \mbox{$ \frac{1}{2}$} F_2^2 {\partial}_2U^i {\partial}_2U^i \;
- \; \mbox{$ \frac{1}{2}$} F_3^2 {\partial}_3U^i {\partial}_3U^i \; \, ,
\end{equation}
in order to account for anisotropies of the crystal lattice. The modified equation for
the thermal propagator,
\begin{equation}
\label{PropEquationAnisotropic}
\Bigg( \frac{{\partial}}{{\partial} x_4} - {\gamma}_1 {\partial}^2_1 
- {\gamma}_2 {\partial}^2_2 - {\gamma}_3 {\partial}^2_3 + \mu H \Bigg) \, G(x)
\, = \, \delta(x) \, ,
\end{equation}
again, allows us to eliminate all terms in the evaluation of graph 8 up to contributions
which involve single space derivatives of the propagator at the origin,
\begin{equation}
\label{z(8)Anisotropic}
z_{8} \; = \; \sum_p \, c_p \, \Big[ {\partial}_p G(x) \Big]_{x=0} \,
\Big[{\partial}_p G(x) \Big]_{x=0} \ , \quad \quad p = 1, 2, 3 \, .
\end{equation}
Even though the thermal propagator is no longer space rotation symmetric, it is still
invariant under parity, such that single space derivatives, evaluated at the origin,
vanish. We conclude that, also for anisotropic lattices, described by the new
contributions (\ref{AnisotropicLagrangian}) in the effective Lagrangian, there is no
interaction term of order $T^3$ in the low-temperature series for the spontaneous
magnetization.

\section{Order parameters of ferro- and antiferromagnets: spontaneous versus staggered
magnetization}
\label{Antiferromagnet}

The low-energy behavior of an O(3) ferromagnet is determined by the spin waves which
represent the Goldstone bosons of this nonrelativistic system. These low-energy
excitations obey a {\it quadratic} dispersion relation. As is well-known, the spin waves
of an {\it anti\/}ferromagnet, on the other hand, follow a {\it linear} dispersion law.
Accordingly, the low-temperature properties of these two systems are quite different --
the difference reveals itself, e.g., in the low-temperature expansion for the
corresponding order parameters. In this section we are going to address the two systems
within the effective Lagrangian perspective.

The low-temperature analysis of an O(N) antiferromagnet, exhibiting a spontaneously
broken internal symmetry O(N) $\to$ O(N-1), was the object of reference.\cite{Hofmann AF}
Here, we only want to point out some basic ingredients in order to compare the system
with the ferromagnet.

In the underlying theory, the O(N) symmetry of the Heisenberg model is explicitly broken
by an external anisotropy field $\vec h$. It is convenient to collect the (N--1)
Goldstone fields $U^a$ in a N-dimensional vector $U^i = (U^0,U^a)$ of unit length,
\begin{equation}
U^i(x) \, U^i(x) \, = \, 1 \, ,
\end{equation}
and to take the constant external field along the zeroth axis, $h^i = (h,0, \dots , 0)$.
The Euclidean form of the effective Lagrangian for an O(N) antiferromagnet up to order
$p^4$ then reads:\cite{Hasenfratz Leutwyler}
\begin{eqnarray}
\label{Leff}
{\cal L}^{AF}_{eff} \ = \ \mbox{$ \frac{1}{2}$} {\cal F}^2 {\partial}_{\mu}
U^i{\partial}_{\mu} U^i \, - \, {\Sigma}_s h^i U^i \, - \, e_1
({\partial}_{\mu} U^i {\partial}_{\mu} U^i)^2 \, - \, e_2 \, ({\partial}_{\mu}
U^i {\partial}_{\nu} U^i)^2 \nonumber\\
+ \, k_1 \! \, \frac{{\Sigma}_s}{{\cal F}^2} \,
(h^i U^i) ({\partial}_{\mu} U^k {\partial}_{\mu} U^k) \, - \, k_2 \, \!
\frac{{\Sigma}_s^2}{{\cal F}^4} \, (h^i U^i)^2 \, - \, k_3 \, \!
\frac{{\Sigma}_s^2}{{\cal F}^4} \, h^i \! h^i \, .
\end{eqnarray}
In the power counting scheme, the field $U(x)$ counts as a quantity of order one.
For the antiferromagnet, derivatives correspond to one power of the momentum,
${\partial}_{\mu} \propto p$, whereas the external field $h$ counts as a quantity of
order $p^2$. Hence, at leading order ($\propto p^2$) two coupling constants, $\cal F$ and
${\Sigma}_s$, occur, at next-to-leading order ($\propto p^4$) we have five such
constants, $e_1, e_2, k_1, k_2, k_3$.

The essential point here is the fact that a term involving a single time derivative does
not show up in the effective Lagrangian: this topological contribution is proportional to
the spontaneous magnetization which, in the case of an antiferromagnet, vanishes. Hence,
the corresponding equation of motion is of second order both in space and in time, its
relativistic structure determining the number of independent magnon states: the Fourier
decomposition contains both positive and negative frequencies, such that a single real
field suffices to describe one particle. Accordingly, there exist {\it two} different
types of spin-wave excitations in an antiferromagnet -- as it is the case in
Lorentz-invariant theories, Goldstone fields and Goldstone particles are in one-to-one
correspondence. These low-energy excitations follow a linear dispersion relation. As a
consequence, in the power counting scheme for antiferromagnetic magnons, powers of
momentum are on the same footing as powers of energy or temperature: $k^2 \propto
{\omega}^2, T^2$.

Let us now turn to the evaluation of the partition function of an O(N) antiferromagnet,
N $\! \geq $2. The relevant graphs are depicted in Figure 4. Lorentz invariance ensures
that only even powers of momentum occur and that loop graphs are suppressed by {\it two}
powers of momentum. Shown are all contributions to the free energy density up to order
$p^6$ or, equivalently, $T^6$. Note the striking difference with respect to Figure 2,
which displays the Feynman graphs relevant for the evaluation of the partition function
of an O(3) ferromagnet.

\begin{figure}[!t]
\centerline{\epsfysize=9.0truecm \hbox{\epsfbox{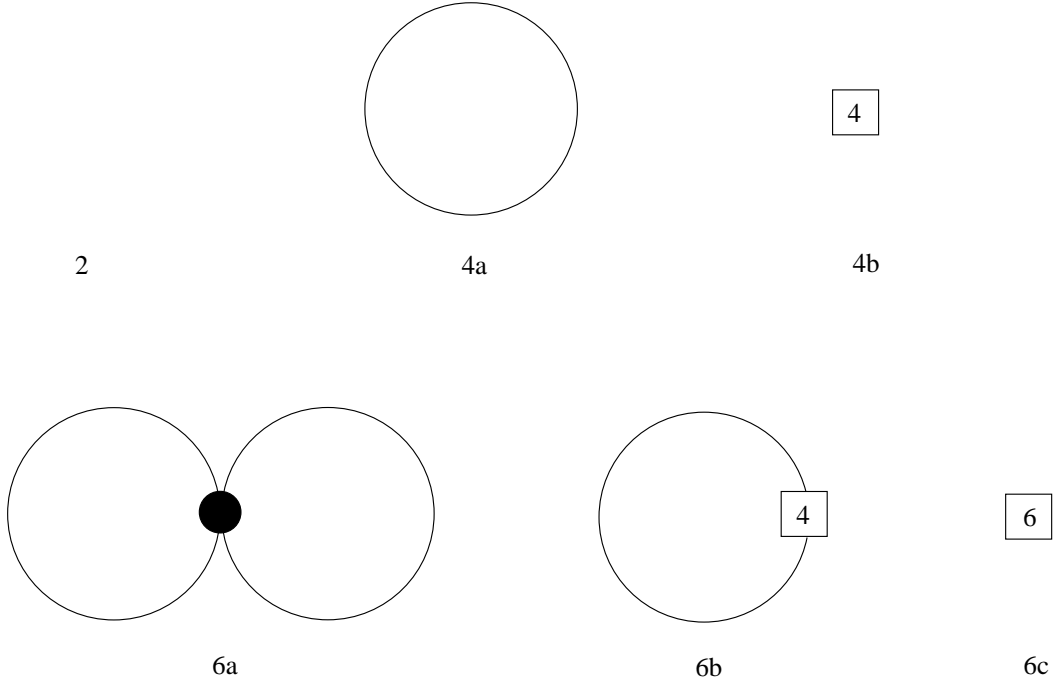}}  }
\medskip
{\tighten \caption[4]{Feynman graphs occurring in the low-temperature expansion for
the partition function of an O(N) antiferromagnet up to order $p^6$. The numbers
attached to the vertices refer to the piece of the effective Lagrangian they come from.
Vertices associated with the leading term of ${\cal L}^{AF}_{eff}$ are denoted by a dot.
The numbers specifying individual graphs correspond to the power of momentum in the
derivative expansion. Note that, in the counting scheme for antiferromagnetic magnons,
two powers of momentum correspond to two powers of energy or temperature.}
\label{fd4} }
\end{figure} 

The tree graphs 4b and 6c are temperature independent and hence merely renormalize the
vacuum energy. One-loop graph 6b exclusively contains a vertex which is quadratic in the
Goldstone boson field and thus only contributes to a renormalization of the mass
$M^2={\Sigma}_s h/{\cal F}^2$ in the free Bose gas term related to graph 4a. The final
expression for the renormalized mass takes the form\cite{Gerber Leutwyler}
\begin{equation}
\label{renMassH}
M_{\pi}^2 \ = \ \frac{{\Sigma}_s h}{{\cal F}^2} \, + \, \frac{N\!-\!3}{32{\pi}^2}
\, \frac{({\Sigma}_s h)^2}{{\cal F}^6} \ln\frac{h}{h_{M}} \; + \; {\cal O} (h^3)
\, .
\end{equation}
The logarithmic scale $h_{M}$ is determined by $k_1$ and $k_2$, i.e., by two
next-to-leading order coupling constants (order $p^4$). The occurrence of such "chiral
logarithms" is characteristic of the effective Lagrangian method in the Lorentz-invariant
regime. To the order $p^6$ considered here, we are thus left with only one candidate for
the interaction: two-loop graph 6a.

In the limit of a zero external field, the low-temperature expansion of the pressure
amounts to
\begin{equation}
\label{PressureAF}
P^{AF}\ = \ \mbox{$ \frac{1}{90}$} {\pi}^2 (N\!-\!1) \, T^4 \, + \, {\cal O} (T^8) \, .
\end{equation}
The $T^4$ contribution represents the free Bose gas term which originates from one-loop
graph 4a. The effective interaction among the Goldstone bosons, remarkably, only
manifests itself through a term of order $T^8$, which is beyond the scope of the present
discussion. As it turns out, the contribution of order $p^6$ coming from two-loop graph
6a, is proportional to the external field, such that, in the limit $h \to 0$ considered
here, it vanishes.

It is interesting to note that this two-loop graph, which involves a four magnon-vertex
from the leading Lagrangian ${\cal L}^2_{eff}$, does not contribute to the pressure of an
O(3) ferromagnet, either. The reason, however, is quite different. For the
antiferromagnet, the contribution is only absent in zero external field, whereas for the
ferromagnet the contribution from graph 8 vanishes because of space rotation symmetry or
parity.

Let us now turn to the order parameters. In the case of an O(3) antiferromagnet this
quantity is referred to as staggered magnetization. It is given by the logarithmic
derivative of the partition function with respect to the external field $h$ and, for an
O(N) antiferromagnet, the low-temperature expansion amounts to (for details
see\cite{Hofmann AF})
\begin{eqnarray}
\label{OrdPar}
{\Sigma}_s(T) \, / \, {\Sigma}_s(0) \ = \ 1 \, - \, \frac{N\!-\!1}{24}
\frac{T^2}{{\cal F}^2} \, 
- \, \frac{(N\!-\!1)\,(N\!-\!3)}{1152} \, \frac{T^4}{{\cal F}^4} \,
+ \, {\cal O} (T^6) \, .
\end{eqnarray}
On the other hand, the low-temperature expansion for the spontaneous magnetization of the
O(3) ferromagnet, derived in the present paper, reads
\begin{displaymath}
\Sigma(T) \, / \, \Sigma(0) \ = \ 1 \, - \, {\alpha}_0 T^{\frac{3}{2}} \,
- \, {\alpha}_1 T^{\frac{5}{2}} \, - \, {\alpha}_2 T^{\frac{7}{2}} \,
- \, {\alpha}_3 T^4 \, + \, {\cal O} (T^{\frac{9}{2}}) \, .
\end{displaymath}
Comparing these formulae for the staggered and spontaneous magnetization, respectively,
the following issues are worth pointing out.

Obviously, as far as the structure of the low-temperature expansion is concerned, the two
series are quite different. The leading terms ($\propto T^2$ for the antiferromagnet,
$\propto T^{3/2}$ for the ferromagnet) are an immediate consequence of the corresponding
dispersion laws: linear for the antiferromagnet, quadratic for the ferromagnet. They are
both related to a one-loop graph.

Now we have to emphasize that the evaluation of the partition function for the
antiferromagnet was performed in a Lorentz-invariant framework, exhibiting a relativistic
dispersion law
\begin{equation}
\label{disprelAFMass}
\omega^2 \, = \, v^4 M^2 + v^2 {\vec k}^2 \, ,
\end{equation}
with the speed of light replaced by the spin-wave velocity $v$. The point is that we did
not consider corrections proportional to ${\vec k}^4, {\vec k}^6, \ldots$ in the
dispersion relation in order to account for the discreteness of the lattice.
Equivalently, we did not consider terms in the subleading pieces ${\cal L}^4_{eff},
{\cal L}^6_{eff}, \ldots $ of the effective Lagrangian which, although space rotation
invariant, are no longer Lorentz invariant. They would have manifested themselves as
terms of order $T^4, T^6, \ldots$ in the staggered magnetization of the antiferromagnet.
Like the analogous terms of order $T^{5/2}, T^{7/2}, \ldots$ in the spontaneous
magnetization of the ferromagnet, these contributions merely correspond to noninteracting
magnons.  

What is remarkable, however, is the fact that the two-loop contribution involving a
four-magnon vertex from the leading piece ${\cal L}^2_{eff}$ of the effective Lagrangian
vanishes in either case: in the low-temperature series for the staggered and spontaneous
magnetization, the spin-wave interaction does not manifest itself through this graph.
Note that the reasons are different. For an O(N) antiferromagnet, there is actually a
term proportional to $T^4$ in the low-temperature series (\ref{OrdPar}) for the order
parameter. But for the particular case N=3 we are considering, the term does not show up
in the staggered magnetization -- corrections associated with the magnon interaction only
appear at order $T^6$.\cite{Hofmann AF} On the other hand, the two-loop contribution for
the ferromagnet, proportional to $T^3$, vanishes due to space rotation symmetry --
corrections originating from the magnon interaction manifest themselves through the Dyson
term proportional to four powers of the temperature.

Another striking difference concerns the fact that for the antiferromagnet, the
low-temperature expansion is completely fixed by a single coupling constant $\cal F$ up
to order $T^4$, i.e., including the first correction coming from the
interaction.\cite{footnote 9} This is not the case for the ferromagnet, where the leading
interaction term, i.e., the two-loop contribution ${\alpha}_3 T^4$, involves coupling
constants from the next-to-leading order Lagrangian ${\cal L}^4_{eff}$. We conclude that,
for  Lorentz-noninvariant systems, kinematics is much less restrictive -- less
information is available via symmetry considerations, such that more phenomenological
input for the corresponding nonrelativistic system is required.

\section{Conclusions and Outlook}
\label{Conclusions}

The low-energy behavior of an O(3) ferromagnet is determined by its low-energy
excitations, the magnons or spin waves, which represent the Goldstone bosons of the
spontaneously broken internal symmetry O(3) $\to$ O(2). The system may be analyzed
within the framework of nonrelativistic effective Lagrangians, which tackles the
phenomenon of spontaneous symmetry breaking from a unified point of view. The method
exploits the symmetry properties of the underlying theory, i.e., the Heisenberg model in
the present case, and formulates the dynamics in terms of Goldstone boson fields. At
large wavelengths, the microscopic structure of condensed matter systems does not play a
significant role: in the corresponding effective Lagrangian, the specific properties of
the system only manifest themselves in the numerical values of a few coupling constants.

The low-energy excitations of an O(3) ferromagnet obey a {\it quadratic} dispersion
relation. In the momentum power counting scheme, which provides us with a systematic
expansion of physical quantities in powers of inverse wavelength, two powers of momentum
thus correspond to only one power of energy or temperature. Remarkably, unlike in a
Lorentz-invariant framework where loop corrections are suppressed by two powers of
momentum, loops involving ferromagnetic magnons are suppressed by three momentum powers. 

In this nonrelativistic framework, the evaluation of the partition function for an O(3)
ferromagnet is presented up to order $p^{10}$: up to order $T^5$ in the free energy
density or, equivalently, up to order $T^4$ in the spontaneous magnetization. In
agreement with Dyson's microscopic analysis, we find that, in the spontaneous
magnetization, the spin-wave interaction only starts contributing at order $T^4$.
Moreover, within the effective Lagrangian perspective we readily understand the absence
of a $T^3$ term in the spontaneous magnetization. This spurious term, which was the
object of various microscopic investigations, vanishes because of space rotation symmetry
or parity of the leading order effective Lagrangian. The effective Lagrangian method not
only proves to be more efficient than the complicated microscopic analysis, but also
addresses the problem from a model-independent point of view based on the symmetries of
the system.

It is very instructive to compare the O(3) ferromagnet and the O(3) antiferromagnet
within the framework of effective Lagrangians. Spin waves in antiferromagnets obey a
linear dispersion relation, implying that powers of momentum, energy and temperature are
all on the same footing: the power counting is identical with the standard,
Lorentz-invariant scheme. Accordingly, the low-temperature properties of these two
systems are quite different as illustrated by the low-temperature expansion for the
pressure and for the corresponding order parameters, the spontaneous and staggered
magnetization, respectively.

Having established the power counting scheme for the ferromagnet, we have paved the way
for further investigations of this nonrelativistic system within the effective Lagrangian
perspective. In fact, the low-temperature expansion for the partition function has
already been carried to order $p^{11}$, where the first three-loop graphs show up. An
outline of the calculation, which goes beyond Dyson's microscopic analysis, will be
presented in a forthcoming paper.\cite{Hofmann in progress}

\acknowledgements
It is a pleasure to thank H. Leutwyler for numerous stimulating discussions and for his
patient assistance throughout this work. Thanks also to S. Mallik, A.V. Manohar, D.
Toublan and J. Soto for their help. I am greatly indebted to the Holderbank-Stiftung for
support. Likewise, support by Schweizerischer Nationalfonds is gratefully acknowledged.


\begin{references}
\bibitem{footnote2}
We will only consider those terms in the higher order pieces of the effective Lagrangian
which are relevant for the evaluation of the partition function. For a more general
analysis the reader is referred to Ref.\cite{Roman Soto}.
\bibitem{footnote3}
The procedure is legitimate, since it merely corresponds to a change of variables.
\bibitem{footnote4}
We neglect contributions due to the anisotropy of the cubic lattice and assume space
rotation symmetry also at order $p^4$ and $p^6$ -- the conclusions of the present work
are not affected by this simplification.
\bibitem{footnote8}
Note that we are now working in Euclidean space.
\bibitem{footnote 9}
In fact, even the coefficient of the interaction term of order $T^6$ is fixed by the
constant $\cal F$ of the leading order Lagrangian ${\cal L}^2_{eff}$ -- next-to-leading
order couplings from ${\cal L}^4_{eff}$ only appear in a logarithmic scale $T_{\Sigma}$
(see Refs.\cite{Hofmann AF,Gerber Leutwyler}).
\end{references}
\end{document}